\documentclass[twocolumn,superscriptaddress,prl,showpacs]{revtex4}
\usepackage{graphicx}
\usepackage{subfigure}
\usepackage{dcolumn}
\usepackage{bm}

\begin{document}
\topmargin 0.0001cm
\title{Search for $\Theta^+(1540)$ pentaquark  in high statistics measurement of 
 $\gamma p \to \bar K^0 K^+ n$  at CLAS\\}


\newcommand*{\INFNGE}{Istituto Nazionale di Fisica Nucleare, Sezione di Genova, and Dipartimento di Fisica, Universit\'a di Genova,
 16146 Genova, Italy}
\affiliation{\INFNGE}
\newcommand*{\RPI}{Rensselaer Polytechnic Institute, Troy, New York 12180-3590}
\affiliation{\RPI}
\newcommand*{\JLAB}{Thomas Jefferson National Accelerator Facility, Newport News, Virginia 23606}
\affiliation{\JLAB}
\newcommand*{\RICE}{Rice University, Houston, Texas 77005-1892}
\affiliation{\RICE}

\newcommand*{\ASU}{Arizona State University, Tempe, Arizona 85287-1504}
\affiliation{\ASU}
\newcommand*{\UCLA}{University of California at Los Angeles, Los Angeles, California  90095-1547}
\affiliation{\UCLA}
\newcommand*{\CSU}{California State University, Dominguez Hills, California  90747-0005}
\affiliation{\CSU}
\newcommand*{\CMU}{Carnegie Mellon University, Pittsburgh, Pennsylvania 15213}
\affiliation{\CMU}
\newcommand*{\CUA}{Catholic University of America, Washington, D.C. 20064}
\affiliation{\CUA}
\newcommand*{\SACLAY}{CEA-Saclay, Service de Physique Nucl\'eaire, F91191 Gif-sur-Yvette, France}
\affiliation{\SACLAY}
\newcommand*{\CNU}{Christopher Newport University, Newport News, Virginia 23606}
\affiliation{\CNU}
\newcommand*{\UCONN}{University of Connecticut, Storrs, Connecticut 06269}
\affiliation{\UCONN}
\newcommand*{\ECOSSEE}{Edinburgh University, Edinburgh EH9 3JZ, United Kingdom}
\affiliation{\ECOSSEE}
\newcommand*{\FIU}{Florida International University, Miami, Florida 33199}
\affiliation{\FIU}
\newcommand*{\FSU}{Florida State University, Tallahassee, Florida 32306}
\affiliation{\FSU}
\newcommand*{\GWU}{The George Washington University, Washington, DC 20052}
\affiliation{\GWU}
\newcommand*{\ECOSSEG}{University of Glasgow, Glasgow G12 8QQ, United Kingdom}
\affiliation{\ECOSSEG}
\newcommand*{\ISU}{Idaho State University, Pocatello, Idaho 83209}
\affiliation{\ISU}
\newcommand*{\INFNFR}{INFN, Laboratori Nazionali di Frascati, Frascati, Italy}
\affiliation{\INFNFR}
\newcommand*{\ORSAY}{Institut de Physique Nucleaire ORSAY, Orsay, France}
\affiliation{\ORSAY}
\newcommand*{\IHEP}{Institute for High Energy Physics, Protvino, 142281, Russia}
\affiliation{\IHEP}
\newcommand*{\ITEP}{Institute of Theoretical and Experimental Physics, Moscow, 117259, Russia}
\affiliation{\ITEP}
\newcommand*{\JMU}{James Madison University, Harrisonburg, Virginia 22807}
\affiliation{\JMU}
\newcommand*{\KHARKOV}{Kharkov Institute of Physics and Technology, Kharkov 61108, Ukraine}
\affiliation{\KHARKOV}
\newcommand*{\KYUNGPOOK}{Kyungpook National University, Daegu 702-701, South Korea}
\affiliation{\KYUNGPOOK}
\newcommand*{\UMASS}{University of Massachusetts, Amherst, Massachusetts  01003}
\affiliation{\UMASS}
\newcommand*{\MOSCOW}{Moscow State University, General Nuclear Physics Institute, 119899 Moscow, Russia}
\affiliation{\MOSCOW}
\newcommand*{\UNH}{University of New Hampshire, Durham, New Hampshire 03824-3568}
\affiliation{\UNH}
\newcommand*{\NSU}{Norfolk State University, Norfolk, Virginia 23504}
\affiliation{\NSU}
\newcommand*{\OHIOU}{Ohio University, Athens, Ohio  45701}
\affiliation{\OHIOU}
\newcommand*{\ODU}{Old Dominion University, Norfolk, Virginia 23529}
\affiliation{\ODU}
\newcommand*{\URICH}{University of Richmond, Richmond, Virginia 23173}
\affiliation{\URICH}
\newcommand*{\SCAROLINA}{University of South Carolina, Columbia, South Carolina 29208}
\affiliation{\SCAROLINA}
\newcommand*{\UNIONC}{Union College, Schenectady, NY 12308}
\affiliation{\UNIONC}
\newcommand*{\VT}{Virginia Polytechnic Institute and State University, Blacksburg, Virginia   24061-0435}
\affiliation{\VT}
\newcommand*{\VIRGINIA}{University of Virginia, Charlottesville, Virginia 22901}
\affiliation{\VIRGINIA}
\newcommand*{\WM}{College of William and Mary, Williamsburg, Virginia 23187-8795}
\affiliation{\WM}
\newcommand*{\YEREVAN}{Yerevan Physics Institute, 375036 Yerevan, Armenia}
\affiliation{\YEREVAN}
\newcommand*{\UK}{University of Kentucky, Lexington, Kentucky 40506}
\affiliation{\UK}
\newcommand*{\UNCW}{University of North Carolina, Wilmington, North Carolina 28403}
\affiliation{\UNCW}
\newcommand*{\UAT}{North Carolina Agricultural and Technical State University, Greensboro, North Carolina 27455}
\affiliation{\UAT}
\newcommand*{\RIKEN}{The Institute of Physical and Chemical Research, RIKEN, Wako, Saitama 351-0198, Japan}
\affiliation{\RIKEN}
\newcommand*{\NOWUNH}{University of New Hampshire, Durham, New Hampshire 03824-3568}
\newcommand*{\NOWUMASS}{University of Massachusetts, Amherst, Massachusetts  01003}
\newcommand*{\NOWMIT}{Massachusetts Institute of Technology, Cambridge, Massachusetts  02139-4307}
\newcommand*{\NOWODU}{Old Dominion University, Norfolk, Virginia 23529}
\newcommand*{\NOWSCAROLINA}{University of South Carolina, Columbia, South Carolina 29208}
\newcommand*{\NOWGEISSEN}{Physikalisches Institut der Universit\"at Gie{\ss}en, 35392 Giessen, Germany}
\newcommand*{\NOWNONE}{unknown, }

\author {M.~Battaglieri} 
\affiliation{\INFNGE}
\author {R.~De~Vita} 
\affiliation{\INFNGE}
\author {V.~Kubarovsky} 
\affiliation{\RPI}
\author {L.~Guo} 
\affiliation{\JLAB}
\author {G.S.~Mutchler} 
\affiliation{\RICE}
\author {P.~Stoler} 
\affiliation{\RPI}
\author {D.P.~Weygand} 
\affiliation{\JLAB}

\author {P.~Ambrozewicz} 
\affiliation{\FIU}
\author {M.~Anghinolfi} 
\affiliation{\INFNGE}
\author {G.~Asryan} 
\affiliation{\YEREVAN}
% ...

\collaboration{The CLAS Collaboration}
     \noaffiliation
%
 
%
%


\date{\today}

\begin{abstract}
The exclusive reaction $\gamma p \to \bar K^0 K^+ n$ was studied 
in the photon energy range between 1.6-3.8 GeV searching for  evidence 
of the exotic baryon $\Theta^+(1540)\to nK^+$. The decay to $nK^+$ requires 
the assignment of strangeness $S=+1$ to any observed resonance.
Data were collected with the 
CLAS detector at the Thomas Jefferson National Accelerator Facility corresponding to an integrated luminosity of 
70 $pb^{-1}$.
No evidence for the $\Theta^+$ pentaquark was found.  Upper limits were
set on the  production cross section as function of center-of-mass angle 
and $nK^+$ mass.
The 95\% CL  upper limit on the total cross section for a narrow resonance 
at 1540 MeV was found to be 0.8 nb.

\end{abstract}
\pacs{12.39.Mk, 13.60.Rj, 13.60.-r, 14.20.Jn, 14.80.-j}

\maketitle
\narrowtext

Following the announcement by  the LEPS 
collaboration~\cite{nakano} in 2003, many 
experiments~\cite{itep,clas,saphir,clas2,itep2,hermes,zeus,serpu,cosytof,jinr} reported
evidence of a  new exotic baryon with strangeness quantum 
number $S=+1$ and valence quark structure $udud\bar{s}$.
The renewed interest in pentaquarks was motivated by a prediction within
the Chiral Soliton Model~\cite{diakonov}  for a $S=+1$ baryon at a mass of 1530 MeV and
width of less than 15 MeV.  
If it exists, this would be the first observation 
of a baryon state that is not made up of a simple 3-quark ($qqq$) valence 
configuration. 
The observation of a second pentaquark, the $\Xi^{--}$ with $dsds\bar{u}$ structure, 
was reported by the NA49 collaboration~\cite{na49} and the  first evidence for an
anti-charmed pentaquark, $\Theta_c$, was found by the H$_1$ Collaboration~\cite{h1}. 
On the other hand, in the past year reanalyses of data collected in high-energy 
experiments~\cite{h1,aleph,babar,belle,bes,cdf,focus,herab,hypercp,lass,l3,phenix,sphinx,wa89} 
show no evidence for pentaquarks, casting doubt on their existence. 
The experimental evidence, both positive and negative, was obtained from data previously 
collected for other purposes in many reaction channels and under very different kinematic conditions, 
which likely involved dissimilar production mechanisms. 
Thus, direct comparisons of the results of the different experiments are very difficult, 
preventing a definitive conclusion about the pentaquark's existence. 
A second generation of dedicated experiments, optimized for the pentaquark search, was undertaken at Jefferson Lab. 
These photoproduction experiments cover the few-GeV beam-energy region where most of the positive evidences were reported, 
with each collecting at least an order of magnitude more statistics than any of the previous measurements. 
The mass resolution  is of the order few MeV and the accuracy of the mass determination is approximately 1-2 MeV, 
allowing precise determination of any possible narrow peaks in the decay distributions.

This Letter presents the first result from the dedicated experimental program at  Jefferson Lab to search for 
pentaquarks. We report on the search for the $\Theta^+$ in  the reaction  $\gamma p \to \bar K^0 K^+ n$.
We searched for the  $\Theta^+$ baryon in its  $K^+ n$ decay mode, which clearly 
identifies the baryonic state in association to the $\bar K^0$ to have positive strangeness.
This channel was previously investigated at ELSA by the SAPHIR collaboration~\cite{saphir} 
in a similar photon energy range, finding positive evidence for a narrow $\Theta^+$ state 
with $M=1540$ MeV and full width half maximum (FWHM) $\Gamma<$ 25 MeV. 
A total production cross section of the order of 300 nb 
(reduced later to 50 nb as reported in Ref.~\cite{saphir50}) was reported. 
For the first time, our new results put previous positive findings to a direct test.\\
\begin{figure}[h]
\vspace{8.cm} 
\includegraphics{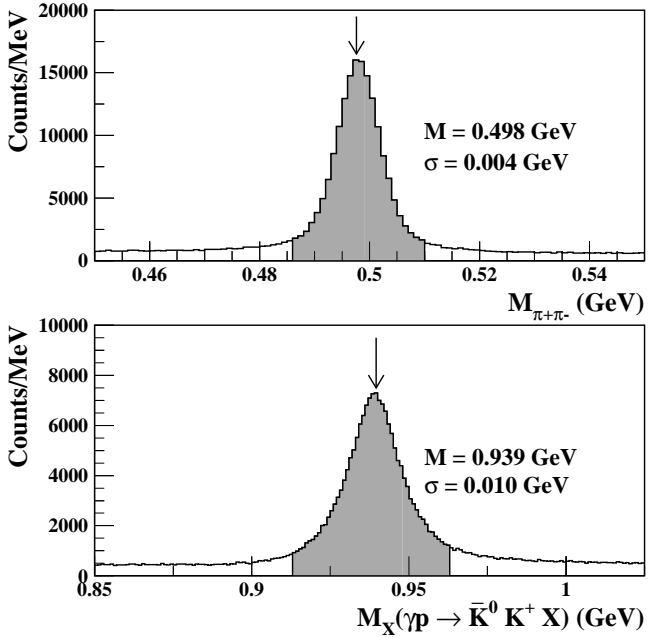}
\caption[]{
Top:  $\pi^+ \pi^-$invariant mass and the  $\bar K^0$  peak. 
Bottom: missing mass for the reaction   $\gamma p \to  \bar K^0 K^+ X$ after $\bar K^0$ selection showing  a peak at  the neutron mass. 
The mass positions and widths of the measured peaks are given. 
For comparison, the arrows indicate the accepted value~\cite{PDG} for the mass position. 
The shaded area corresponds to the events used in the analysis.}
\label{fig1}
\end{figure}
This measurement was performed using the CLAS~\cite{B00} detector at Jefferson Lab in the experimental Hall-B
 with a bremsstrahlung photon beam produced by a primary continuous electron beam of energy $E_0$ = 4.0 GeV. 
A bremsstrahlung tagging system~\cite{SO99}, which measures the energy of each interacting photon with resolution of 0.1$\%$ $E_0$ 
was used to tag photons in the energy range 1.6-3.8 GeV. 
The target consisted of a 40-cm-long cylindrical cell containing liquid hydrogen. 
Outgoing hadrons were detected and identified in CLAS.
Momentum information for charged particles was obtained via tracking
through three regions of multi-wire drift chambers~\cite{DC} inside  a toroidal magnetic 
field ($\sim 0.5$ T), which was generated by six superconducting coils. 
The CLAS momentum resolution is of the order of 0.5-1\% ({\bf $\sigma$}) depending on
the kinematics. The detector geometrical acceptance for each positive particle in the 
relevant kinematic region is about 40\%. It is somewhat less for low-energy negative 
hadrons, which can be lost at  forward angles because they are bent out of the acceptance
by the toroidal field. The field was set to bend the positive particles away from the beam into the acceptance region
of the  detector. Time-of-flight scintillators (TOF) were used for hadron
identification~\cite{Sm99}. 
The interaction time between the incoming photon and the target
was measured by the Start Counter (ST)~\cite{ST}, consisting of a set of 24 2.2 mm thick plastic scintillators 
surrounding the hydrogen cell.
Coincidences between the photon tagger and two charged particles in the CLAS detector triggered the recording of the events. 
An integrated luminosity of about 70 pb$^{-1}$  was accumulated in 50 days of running. 
In total, about 20 TB of data were collected.

\begin{figure}[h]
\vspace{8.cm} 
\includegraphics{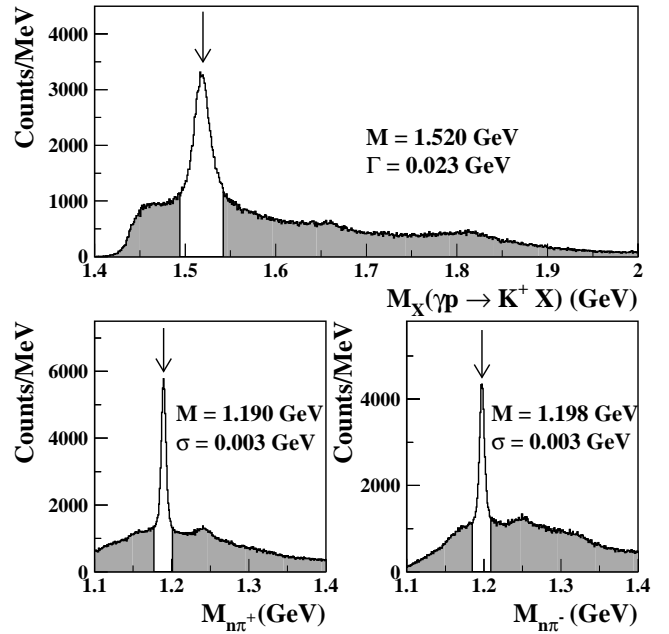}
\caption[]{
Top: $K^+$ missing mass distribution with the $\Lambda^*(1520)$ peak. 
Bottom:  $n\pi^+$ (left) and $n\pi^-$ (right) invariant mass distributions 
with $\Sigma^+(1189)$ and $\Sigma^-(1197)$ peaks. 
The mass position and width of the measured peaks are indicated. 
For comparison, the arrows indicate the accepted value~\cite{PDG} for the mass position. 
The shaded area corresponds to the events used in the analysis.}
\label{fig2}
\end{figure}

The reaction $\gamma p \to \bar K^0 K^+ n$ was isolated as follows. 
The $K^+$ was detected directly in the magnetic spectrometer, and the $K_S$ component of the $\bar K^0$
was reconstructed from its  $\pi^+\pi^-$ decay. 
The momentum  of the neutron was reconstructed from the known incident photon energy 
and measurements of all other particles in the event. 
Calibrations of all detector components, and especially the tagger system, were performed 
achieving a precision of 1-2 MeV in the $nK^+$ invariant mass determination. 
The quality of the channel identification is shown in Fig.~\ref{fig1}  
where the  $\bar K^0$  and the missing neutron peaks are seen above a small background.

\begin{figure}[h]
\vspace{8.cm} 
\includegraphics{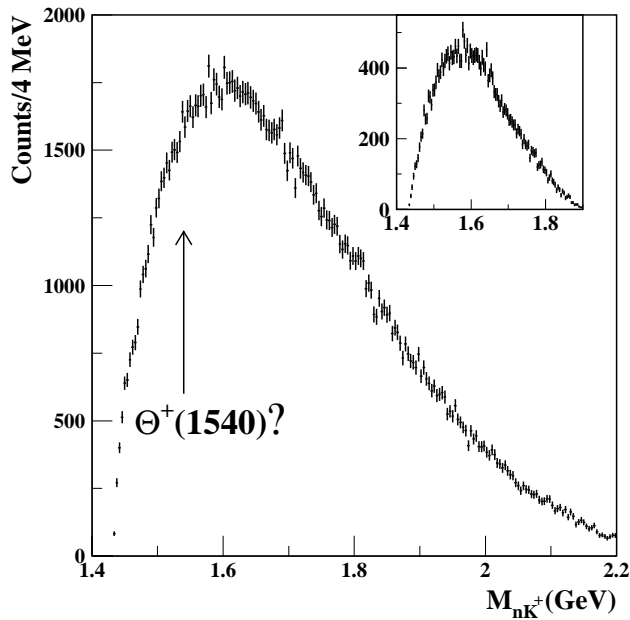}
\caption[]{
The $nK^+$ invariant mass distribution after all cuts. 
It is smooth and no narrow structures are evident. 
The arrow shows the position where evidence for the $\Theta^+$ was found by previous experiments. 
The inset shows the $nK^+$ mass distribution with specific cuts to reproduce the SAPHIR analysis~\cite{saphir} as described in the text.}
\label{fig3}
\end{figure}

Reactions involving the decays of hyperons  also contribute to the same final state.
 The most significant are: $\gamma p  \to  K^+ \Lambda^*(1520) \to K^+ \bar K^0 n$,
$\gamma p  \rightarrow  \pi^-  K^+ \Sigma^+$, and $\gamma p  \rightarrow  \pi^+ K^+ \Sigma^-$.
These reactions are backgrounds 
to the pentaquark search, but are easily removable in our analysis with cuts around the known masses. 
They also serve  as checks of our analysis procedure, e.g. by comparing  their production cross sections  with the world data.
Figure~\ref{fig2}  shows the background hyperon peaks: $\Lambda^*(1520)$ in the  $K^+$ missing mass spectrum
and the $\Sigma^+$,  $\Sigma^-$ peaks in the $n\pi^+$  and $n\pi^-$
invariant mass spectra respectively. The mass region of each of these hyperon peaks was excluded from the final data set.
After all cuts, the data sample contained approximately $0.17\times10^6$ events 
out of the $7\times10^9$ in the original data set. 
The resulting  $nK^+$  invariant mass distribution is shown in Fig.~\ref{fig3}.
The spectrum is smooth and structureless. 
In particular, no evidence for a peak or an enhancement is observed at masses near 1540 MeV, 
where signals associated with the $\Theta^+$ were previously reported.

To enhance our sensitivity to a possible resonance signal not visible in the integrated distribution, 
we considered  the  two-body reaction $\gamma p \to \bar K^0 \Theta^+(1540)$
and selected different $K_S$ ($\bar K^0$) center-of-mass angle intervals.
Monte Carlo studies of the CLAS acceptance for this reaction showed that we could detect 
events over the entire angular range (0$^\circ$-180$^\circ$), 
with some reduction of efficiency at forward angles ($\theta^{CM}_{\bar K^0}<30^\circ$). 
No structures were found in the distribution when specific angular ranges were selected. 

Since no signal was found, an upper limit for the $\Theta^+$  production cross section in this reaction channel was extracted.
The unbinned $nK^+$  mass spectrum was fit in the range 1.45 - 1.8 GeV using a maximum likelihood procedure, 
with the sum of a narrow Gaussian function 
and a $5^{th}$-order polynomial that parameterizes, respectively, the $\Theta^+$ contribution and a smooth background.
To derive the corresponding event yields, the fitted functions were integrated over $\pm 3 \sigma$ around a fixed mass position.
The fit procedure  was repeated varying the resonance position from 1520 to 1600 MeV in 5 MeV steps while the width $\sigma$  was fixed at  3.5 MeV.
This value was  derived by Monte Carlo simulation assuming a negligible intrinsic width as suggested from recent analyses
of $KN$ scattering data~\cite{kn} and therefore dominated by the CLAS experimental resolution. 
The validity of the Monte Carlo simulations in reproducing the experimental data was checked by comparing the predicted 
with measured widths of narrow states such as the $\Sigma^+$ and  $\Sigma^-$.

The data set was independently analyzed by three groups, each one deriving the estimate of the $\Theta^+$ and the 
background yields. The three analyses differ in the reaction selection cuts, in the  background rejection criteria,  
and in the fit of the mass spectra.
The three results were found consistent and  combined together taking the average of the event yields, for both signal and background,
in the conservative assumption of totally correlated measurements. These values were then used to evaluate an upper limit 
at 95\% CL on the $\Theta^+$  yield using the Feldman and Cousins approach~\cite{fc99}.

The upper limit on the yields was then transformed into an upper limit on the $\Theta^+$
production cross section taking into account the luminosity of incident photons and target,
the CLAS detection acceptance, the $\bar K^0 \to K_S\to \pi^+ \pi^-$ branching ratios of 50\%$\times$69\%~\cite{PDG},
the assumed $\Theta^+$  branching ratio to $nK^+$  of 50\%,
and several models for the production mechanism.
The CLAS acceptance for the detection of the $\Theta^+$ in this reaction was obtained by means of detailed Monte Carlo 
studies which included knowledge of the detector geometry and response to traversing particles. 
In the simulation the $\gamma p \to \bar K^0 \Theta^+ \to \pi^+\pi^- K^+ n$ distributions were generated assuming five
different $\Theta^+$ production mechanisms: $t$-exchange dominance (the $\bar K^0$ is mainly produced at forward angles 
in the center-of-mass system), $u$-exchange dominance (at backward angles),  uniformly distributed, 
and using the predictions of the model in Ref~\cite{oh04} (with and without $K^*$ exchange process).
For the $t$-exchange hypothesis we used the same angular distribution as for  $\gamma p \to  K^+ \Lambda^*(1520)$ production, 
which exhibits a typical $t$-channel forward peaking behavior~\cite{lstar-nina}. 
The $u$-exchange distribution was generated the same way but interchanging the center-of-mass angles of the  $\bar K^0$ and $\Theta^+$.
The CLAS overall detection efficiencies obtained with different production mechanisms 
varied between 2.8\% for the $t$-exchange hypothesis and 5.2\% for the angular distribution of Ref.~\cite{oh04} when no $K^*$ 
exchange process is included. All the upper limits reported in this article  were derived in the most conservative scenario, i.e. 
in the $t$-exchange hypothesis.

\begin{figure}[h]
\vspace{8.cm} 
\includegraphics{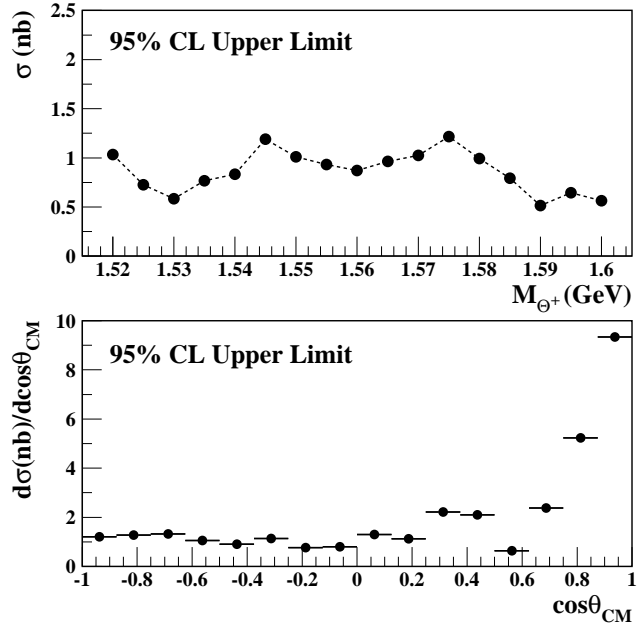}
\caption[]{
The 95\% CL upper limit on the total cross section as a function of the  $\Theta^+$ mass (top) 
and on the differential cross section $d\sigma/d\cos\theta^{\bar K^0}_{CM}$ (bottom) for the reaction   $\gamma p \to \bar K^0 \Theta^+$ for an assumed
 $\Theta^+$ mass of 1540 MeV. The dotted line in the top plot is to guide the eye.
}
\label{fig5}
\end{figure}

The upper panel in Fig.~\ref{fig5} shows the upper limit on the total cross section as a function of the $\Theta^+$ mass.
An upper limit of 0.8 nb was found for $M=1540$ MeV. 
The process to extract the yield was repeated for each angular bin to derive the 
95\% CL upper limit on the  $\Theta^+(1540)$ differential cross section  $d\sigma/d\cos\theta^{\bar K^0}_{CM}$. 
The result is shown in the lower panel of Fig.~\ref{fig5}. 
The cross section upper limit remains within about 1-2 nb for most of the angular range and 
rises at forward angles due to the reduced CLAS acceptance. 
As a check on our procedure, we extracted  the differential and the total cross section for several known reactions from the same data set, 
finding overall good agreement within experimental uncertainties with the existing world measurements. 
These results will be reported elsewhere. 

Another measure of the strength of the pentaquark signal is to compare the upper limit on the yield to the number of $\Lambda^*(1520)$
events produced in the reaction. The  95\% CL upper limit on the number of $\Theta^+$ events at a mass of 1540 MeV for our data
sample is 220 events. The number of  observed  $\Lambda^*(1520)$ events, shown in the upper panel of Fig.~\ref{fig2}, was
determined using a Breit-Wigner resonance shape fit to be 100k. Thus, the ratio  is less than 220/100k = 0.22 \%  (95\% CL). 

Our upper limit on the cross section is in clear disagreement with the findings of Ref.~\cite{saphir} which reported a $\Theta^+$
signal of 63 events at a mass of 1540 MeV corresponding to the published total cross section of 300 nb.
In order to better compare with that experiment, we repeated the analysis applying the same cuts reported in that paper: 
the photon energy was limited to 2.6 GeV, only events with a forward-emitted $\bar K^0$ ($\theta^{\bar K^0}_{CM}> 60^\circ$) 
were used and no cuts were made to exclude  hyperons. 
The resulting mass distribution is shown in the inset of Fig.~\ref{fig3}: it remains smooth and structureless. 
Another way to show the inconsistency of the two experiments is to compare the ratio of the upper limit of the number of $\Theta^+$
 with the number of the observed $\Lambda^*(1520)$. 
Applying again the same specific cuts to reproduce the SAPHIR analysis, we evaluated a 95\% CL limit on the $\Theta^+$
 yield of less than 100 events. In the same photon energy range (1.6-2.6 GeV) we observed $\sim 53000$ $\Lambda^*(1520)$'s,
 compared with a  $\Theta^+$ yield of 63 and 630 $\Lambda^*(1520)$'s respectively reported in Ref.~\cite{saphir}. 
The ratios obtained in the two experiments differ by more than a factor 50.

In conclusion, this is the first result of a dedicated set of high-statistics and high-resolution experiments 
undertaken at Jefferson Lab to elucidate the debate on the existence of the pentaquark. 
The reaction $\gamma p \to \bar K^0 K^+ n$  was studied in search for evidence of the $\Theta^+$  pentaquark in the 
 $nK^+$ decay channel. 
The final state was isolated detecting the  $K^+$, the $\bar K^0$ by its  $\pi^+\pi^-$ decay, and identifying the neutron by means of the missing mass technique. 
The direct measurement of the  $K^+$  allows one to define the strangeness of any  baryon resonance observed in this final state. 
The $nK^+$ mass distribution was found to be smooth and structureless. No evidence for a narrow resonance was found in the mass range 1520-1600 MeV.
An upper limit of 0.8 nb (95\% Confidence Level) on the total production cross section for a $\Theta^+$ mass of 1540 MeV was set. 
This is in disagreement with previously reported evidence for a resonance in the same reaction channel,
and sets stringent upper-limits on the models which predict these long-lived pentaquark states.

We would like to acknowledge the outstanding efforts of the staff of the Accelerator
and the Physics Divisions at Jefferson Lab that made this experiment possible. 
This work was supported in part by  the  Italian Istituto Nazionale di Fisica Nucleare, 
the French Centre National de la Recherche Scientifique
and Commissariat \`a l'Energie Atomique, 
the U.S. Department of Energy and National Science Foundation, 
and the Korea Science and Engineering Foundation.
The Southeastern Universities Research Association (SURA) operates the
Thomas Jefferson National Accelerator Facility for the United States
Department of Energy under contract DE-AC05-84ER40150.


\begin{thebibliography}{99}


\bibitem{nakano}  T. Nakano {\it et al.} (LEPS Collaboration), Phys. Rev. Lett. {\bf 91}, 012002 (2003). 
\bibitem{itep} V.~V.~Barmin {\it et al.} (DIANA Collaboration), Phys.\ Atom.\ Nucl.\  {\bf 66}, 1715 (2003). 
\bibitem{clas} S. Stepanyan {\it et al.} (CLAS Collaboration), Phys. Rev. Lett. {\bf 91}, 252001 (2003).
\bibitem{saphir} J. Barth {\it et al.} (SAPHIR Collaboration),   Phys. Lett. B {\bf 572}, 127 (2003). 
\bibitem{clas2} V. Kubarovsky {\it et al.} (CLAS Collaboration), Phys. Rev. Lett. {\bf 92}, 032001 (2004).
\bibitem{itep2} A.E. Asratyan, A.G. Dolgolenko, and M.A. Kubantsev, Phys.\ Atom.\ Nucl.\  {\bf 67}, 682 (2004). 
\bibitem{hermes} A. Airapetian {\it et al.} (HERMES Collaboration), Phys.\ Lett.\ B {\bf 585}, 213 (2004).
\bibitem{zeus} S.~Chekanov {\it et al.} (ZEUS Collaboration), Phys.\ Lett.\ B {\bf 591}, 7 (2004).
\bibitem{serpu} A.~Aleev {\it et al.} (SVD Collaboration), Phys. At. Nucl. 68, 974 (2005) and hep-ex/0509033.
\bibitem{cosytof} M.~Abdel-Bary {\it et al.} (COSY-TOF Collaboration), Phys.\ Lett.\ B {\bf 595}, 127 (2004).
\bibitem{jinr} P.~Z.~Aslanyan, V.~N.~Emelyanenko and G.~G.~Rikhkvitzkaya, arXiv:hep-ex/0403044.
\bibitem{diakonov} D. Diakonov, V. Petrov and M. Polyakov, Z. Phys. A {\bf 359}, 305 (1997).  

\bibitem{na49} C. Alt {\it et al.} (NA49 Collaboration), Phys.\ Rev.\ Lett.\  {\bf 92}, 042003 (2004).
\bibitem{h1} A.~Aktas {\it et al.} (H1 Collaboration), Phys.\ Lett.\ B {\bf 588}, 17 (2004).



\bibitem{aleph} S.~Schael {\it et al.} (ALEPH Collaboration), Phys.\ Lett.\ B {\bf 599}, 1 (2004).
\bibitem{babar} B.~Aubert {\it et al.} (BABAR Collaboration), arXiv:hep-ex/0408064.
\bibitem{belle} K.~Abe {\it et al.} (Belle Collaboration]), arXiv:hep-ex/0411005.
\bibitem{bes} J.~Z.~Bai {\it et al.} (BES Collaboration), Phys.\ Rev.\ D {\bf 70}, 012004 (2004).
\bibitem{cdf} I.~V.~Gorelov (CDF Collaboration), arXiv:hep-ex/0408025; D.~O.~Litvintsev (CDF Collaboration), Nucl.\ Phys.\ Proc.\ Suppl.\  {\bf 142}, 374 (2005).
\bibitem{focus} K.~Stenson (FOCUS Collaboration), arXiv:hep-ex/0412021.
\bibitem{herab} I.~Abt {\it et al.} (HERA-B Collaboration), Phys.\ Rev.\ Lett.\  {\bf 93}, 212003 (2004); K.~T.~Knopfle, M.~Zavertyaev and T.~Zivko (HERA-B Collaboration),
 J.\ Phys.\ G {\bf 30}, S1363 (2004).
\bibitem{hypercp} M.~J.~Longo {\it et al.} (HyperCP Collaboration), Phys.\ Rev.\ D {\bf 70}, 111101(R) (2004).
\bibitem{lass} J.~Napolitano, J.~Cummings and M.~Witkowski, arXiv:hep-ex/0412031.
\bibitem{l3} S.~R.~Armstrong, Nucl.\ Phys.\ Proc.\ Suppl.\  {\bf 142}, 364 (2005).
\bibitem{phenix} C.~Pinkenburg (PHENIX Collaboration), J.\ Phys.\ G {\bf 30}, S1201 (2004).
\bibitem{sphinx} Y.~M.~Antipov {\it et al.} (SPHINX Collaboration), Eur.\ Phys.\ J.\ A {\bf 21}, 455 (2004).
\bibitem{wa89} M.~I.~Adamovich {\it et al.} (WA89 Collaboration), arXiv:hep-ex/0405042.
 

\bibitem{saphir50} M. Ostrick, Prog. Part. Nucl. Phys. {\bf 55}, 337 (2005).
\bibitem {B00} B. Mecking {\it et al.}, Nucl. Instr. and Meth. {\bf A503}, 513  (2003).
\bibitem{SO99}  D.I. Sober {\it et al.}, Nucl. Instr. and Meth. {\bf A440}, 263  (2000).
\bibitem{DC}    M.D. Mestayer    {\it et al.}, Nucl. Instr. and Meth. {\bf A449}, 81 (2000).
\bibitem{Sm99}  E.S. Smith {\it et al.}, Nucl. Instr. and Meth. {\bf A432}, 265  (1999).
\bibitem{ST} G. Mutchler {\it et al.}, submitted to  Nucl. Instr. and Meth.
\bibitem{PDG}  S. Eidelman {\it et al.}, Phys. Lett. {\bf B592}, 1 (2004). 
 \bibitem{kn} R. A.~Arndt, I.~I.~Strakovsky, R.~L.~Workman, Phys.\ Rev.\ {\bf C68}, 042201(R) (2003); 
R.~N. Cahn and G.~H.~Trilling, Phys. Rev. {\bf D69}, 011501(R) (2004);  J. Haidenbauer and G. Krein,  Phys. Rev. {\bf C68}, 
052201(R) (2003);  W.~R.~Gibbs,  Phys. Rev. {\bf C70},  045208 (2004); A. Sibirtsev {\it et al.}, Phys.\ Lett.\  {\bf B599}, 230 (2004).
\bibitem{fc99} G.~J.~Feldman and R.~D.~Cousins, Phys.\ Rev.\ D {\bf 57} 3873 (1998).
\bibitem{oh04} Y.~Oh, H.~Kim and S.~H.~Lee, Phys.\ Rev.\ D {\bf 69} (2004) 014009.
\bibitem{lstar-nina}  D. Barber  {\it et al.}, Z. Phys. {\bf C7}, 17  (1980). 


\end{thebibliography}
\end{document}